\documentclass[runningheads]{llncs}
\usepackage{graphicx}
\usepackage{array}
\usepackage[ruled]{algorithm2e}
\usepackage{dsfont}
\usepackage{amsmath}
\usepackage{amsfonts,amssymb}
\usepackage{booktabs}
\usepackage{bm}
\usepackage{stfloats}
\usepackage{multirow}
\usepackage{hyperref}
\usepackage{float}
\usepackage{xcolor}
\usepackage{enumerate}
\usepackage[misc]{ifsym}

\newcommand\eg{\emph{e.g.},\xspace}

\newcommand\figref[1]{Figure~\ref{#1}}
\newcommand\wrt{\emph{w.r.t.}\xspace}
\newcommand\equref[1]{Eq.~(\ref{#1})}

\newcommand\ie{\emph{i.e.},\xspace}
\newcommand\algoref[1]{Algorithm~\ref{#1}}
\newcommand\tabref[1]{Table~\ref{#1}}

\begin{document}
%
\title{Chain-of-Choice Hierarchical Policy Learning for Conversational Recommendation}

\titlerunning{Chain-of-Choice Hierarchical Policy Learning}
\author{Wei Fan\inst{1} \and
Weijia Zhang\inst{2} \and
Weiqi Wang\inst{1} \and
Yangqiu Song\inst{1} \and
Hao Liu\inst{2}\textsuperscript{(\Letter)}
}
\authorrunning{W. Fan et al.}

\institute{The Hong Kong University of Science and Technology, Hong Kong, China\\
\email{wfanag@connect.ust.hk, \{wwangbw, yqsong\}@cse.ust.hk}\and
The Hong Kong University of Science and Technology (Guangzhou),\\ Guangzhou, China\\
\email{wzhang411@connect.hkust-gz.edu.cn, liuh@ust.hk}}
\maketitle              
\begin{abstract}
Conversational Recommender Systems (CRS) illuminate user preferences via multi-round interactive dialogues, ultimately navigating towards precise and satisfactory recommendations.
However, contemporary CRS are limited to inquiring binary or multi-choice questions based on a single attribute type~(\eg color) per round, which causes excessive rounds of interaction and diminishes the user's experience.
To address this, we propose a more realistic and efficient conversational recommendation problem setting, called \textbf{M}ulti-\textbf{T}ype-\textbf{A}ttribute \textbf{M}ulti-round \textbf{C}onversational \textbf{R}ecommendation~(\textbf{MTAMCR}), which enables CRS to inquire about multi-choice questions covering multiple types of attributes in each round, thereby improving interactive efficiency.
Moreover, by formulating MTAMCR as a hierarchical reinforcement learning task, we propose a \textbf{C}hain-\textbf{o}f-\textbf{C}hoice \textbf{H}ierarchical \textbf{P}olicy \textbf{L}earning (\textbf{CoCHPL}) framework to enhance both the questioning efficiency and recommendation effectiveness in MTAMCR. 
Specifically, a long-term policy over options~(\ie ask or recommend) determines the action type, while two short-term intra-option policies sequentially generate the chain of attributes or items through multi-step reasoning and selection, optimizing the diversity and interdependence of questioning attributes.
Finally, extensive experiments on four benchmarks demonstrate the superior performance of CoCHPL over prevailing state-of-the-art methods.

\keywords{Conversational Recommendation \and Hierarchical Reinforcement Learning \and Graph Representation Learning}
\end{abstract}

\begin{figure}[ht]
	
	\begin{minipage}{0.32\linewidth}
		\centerline{\includegraphics[width=\textwidth]{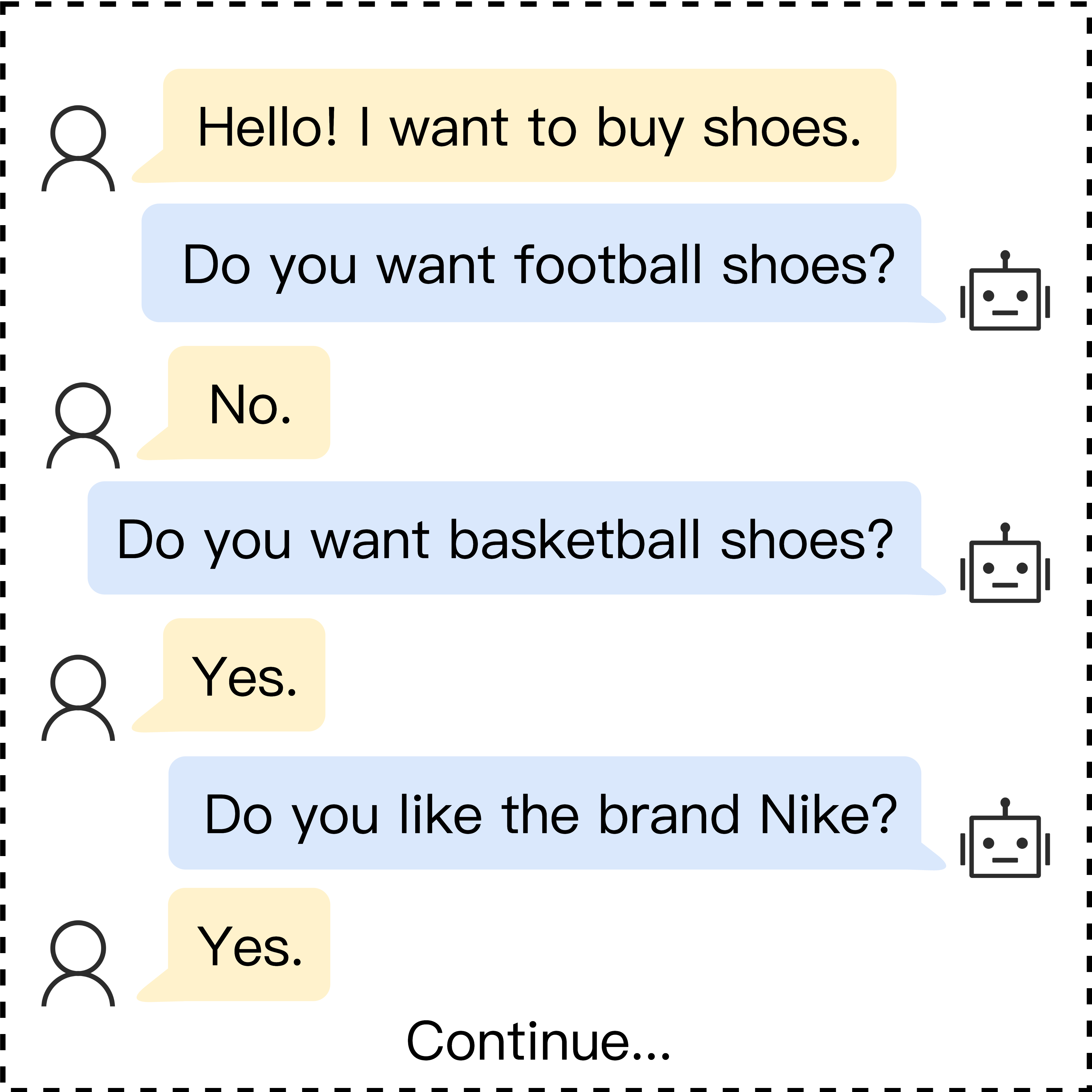}}
		\centerline{(a) Binary choices}
	\end{minipage}
	\begin{minipage}{0.32\linewidth}
		\centerline{\includegraphics[width=\textwidth]{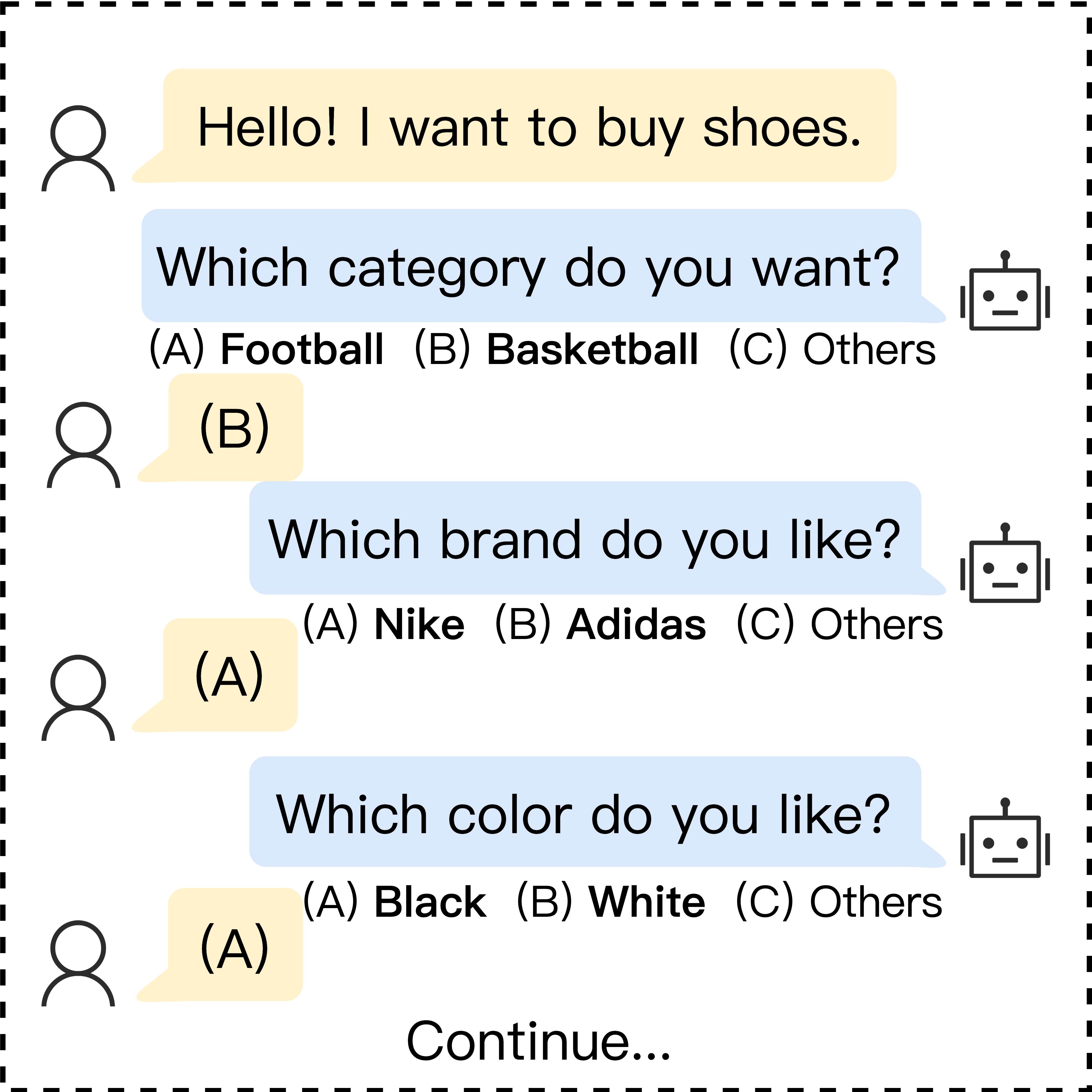}}
		\centerline{(b) Single-Type attributes}
	\end{minipage}
	\begin{minipage}{0.32\linewidth}
		\centerline{\includegraphics[width=\textwidth]{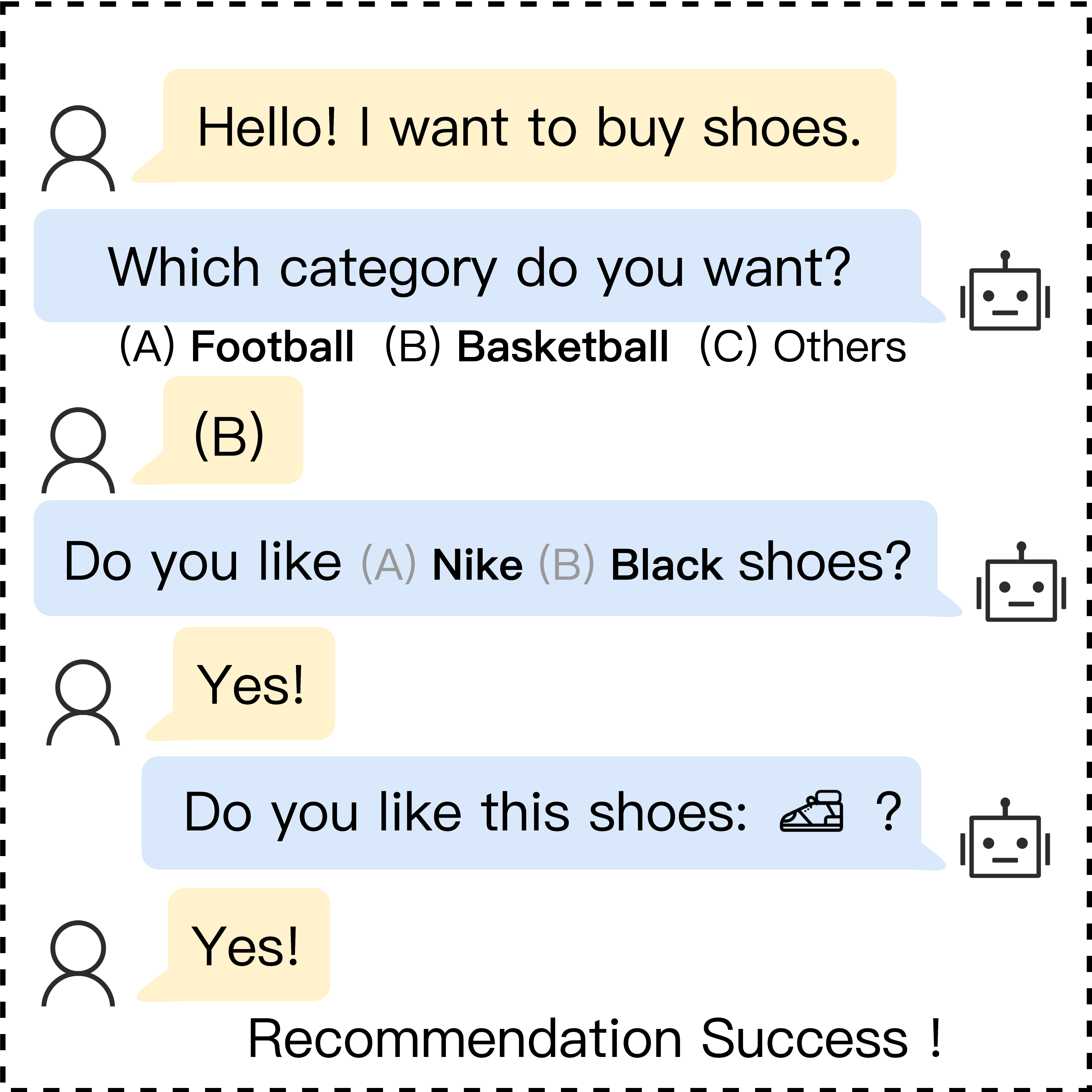}}
		\centerline{(c) Multi-Type attributes}
	\end{minipage}
 
	\caption{Example of different conversational recommendation settings.}
	\label{fig:intro}
\end{figure}
 
\section{Introduction}
Compared to traditional recommender systems, conversational recommender system (CRS)~~\cite{sun2018conversational} serves as an online salesperson who elicits user preferences~\cite{xie2021comparison,lin2023enhancing} through engaging question-and-answer dialogues~~\cite{lei2020estimation,zhang2018towards} to provide tailored recommendations.
This conversational approach allows for explicit preference reasoning by actively asking questions about attributes and making desirable recommendations based on the user's real-time feedback.

Recent research has explored different settings to create more realistic conversational recommendation scenarios.
Existing studies allow CRS to ask either binary (yes/no) questions for each selected attribute~\cite{sun2018conversational,lei2020estimation,lei2020interactive,deng2021unified} or multiple choice questions falling into one selected attribute type (\eg brand, color) in each round~\cite{zhang2022multiple}.
However, existing CRS using binary choices~\cite{deng2021unified} or single-type attribute questions~\cite{zhang2022multiple} can be repetitive and time-consuming, causing frustration for the user and decreasing the overall recommendation success rate. Consider an example illustrated in \figref{fig:intro}, where the user wants shoes with the attributes ``Basketball,'' ``Nike,'' and ``Black.'' In \figref{fig:intro}(a) and \figref{fig:intro}(b), the question choices are restricted to one attribute type per round, necessitating at least three rounds to extract all the preferred attributes. Our work studies a more realistic scenario, namely Multi-Type-Attribute Multi-round Conversational Recommendation~(MTAMCR) as depicted in \figref{fig:intro}(c), which enables CRS to generate questions encompassing diverse yet dependent attribute types. By optimizing the combination of attribute types, we can enhance the efficiency of reasoning and questioning, ultimately increasing the success rate of recommendations.

Previous works~\cite{sun2018conversational,lei2020estimation,lei2020interactive,deng2021unified,zhang2022multiple,zhao2023towards} have formulated conversational recommendation as a Markov Decision Process~(MDP) and introduced reinforcement learning~(RL) to choose the attributes and recommended items. However, these approaches encounter two main limitations when apply them to the new MTAMCR scenario:
1) Diversity of attributes: Previous works select the attribute instances within a single attribute type, allowing for exploring only one type of user's attribute preference~(\eg color) per round. Thus, these methods need more rounds to discover and capture all the attributes of the user's target item.
2) Dependency between attributes: Existing RL-based approaches directly select one or multiple attribute instances based on the top-K Q-values without considering the influence of previous attribute choices on subsequent attribute selections.
As a result, such approaches usually neglect the dependency between selections, leading to suboptimal combinations of attributes.

\begin{figure}[t]
	\centering
        \begin{minipage}{0.65\linewidth}
 	    \centerline{\includegraphics[width=\textwidth]{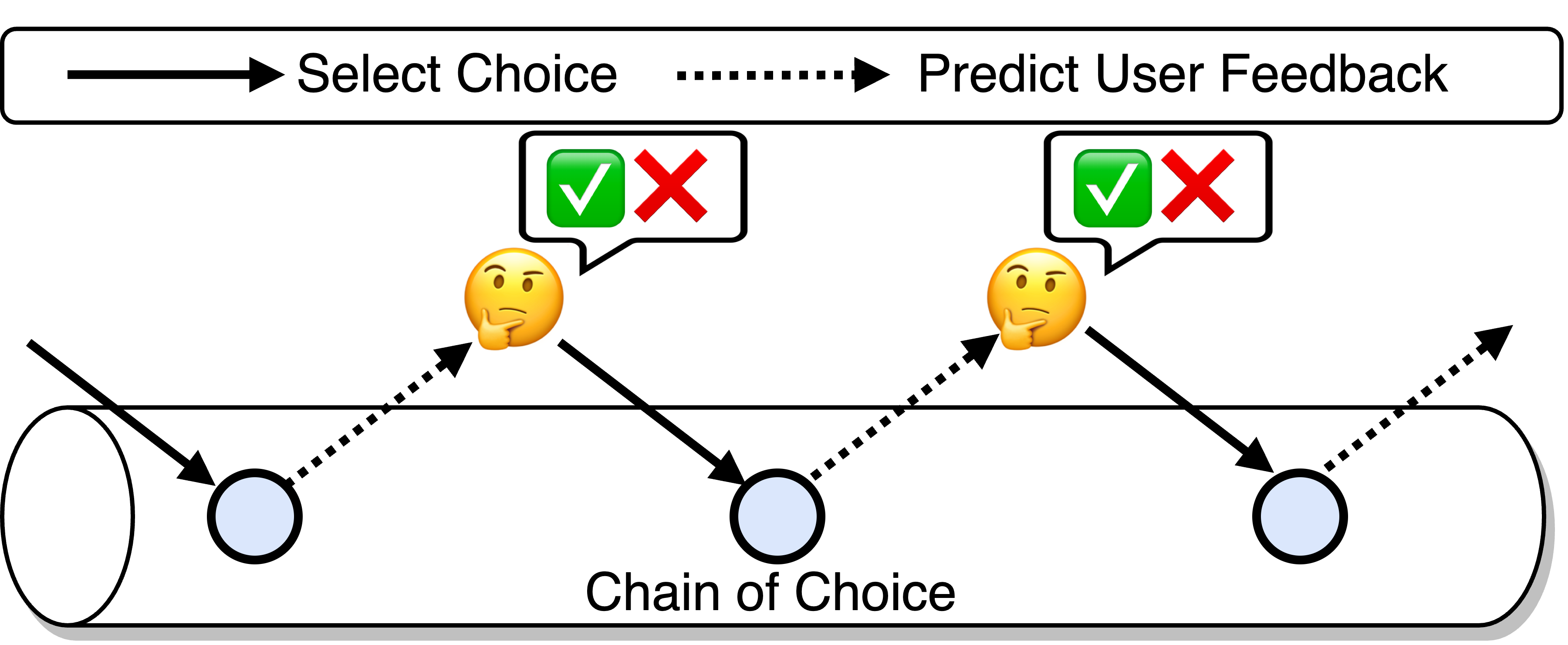}}
        \end{minipage}
 
	\caption{During each turn, the agent engages in a decision-making process where it selects a choice and then predicts the user's feedback in order to infer the subsequent state. With the predicted state, the agent selects the next choice, continually repeating this process until the round eventually reaches its termination point.}
	\label{fig:chain}
\end{figure}

To address the two limitations, we propose the chain-of-choice generation approach, as illustrated in \figref{fig:chain}. This approach infers the next choice, considering the previous choices by predicting the user feedback. For example, if the user wants basketball shoes and the first choice in a given turn is ``Nike,'' selecting the color ``Black'' as the next choice is more effective than presenting a brand selection between ``Nike'' and ``Adidas.'' This is because we predict that the user would accept the attribute ``Nike'' since many individuals highly prefer Nike basketball shoes. Therefore, it is strategically advantageous to prioritize exploring color preferences rather than spending more time deliberating over brand preferences. With the chain of choices, we first proposed the Chain-of-Choice Hierarchical Policy Learning~(CoCHPL) framework, which formulates MTAMCR as a hierarchical reinforcement learning task. In each turn, CoCHPL employs a long-term policy to select the option~(\ie ask or recommend) and generates the chain of choices~(\ie attributes or items) step by step via different short intra-option policies. To model the transitions of the choice chain and determine the optimal moment for terminating the current choice generation, we utilize learnable functions that predict feedback for the ongoing choice chain and infer the next state for the next choice selection.

Our major contributions can be summarized as follows:
\begin{enumerate}[1)]
    \item We introduce the \textbf{MTAMCR} scenario, which allows the CRS agent to ask questions under the same or different attribute types in each round, enabling more realistic and effective user interaction.
    \item We propose the \textbf{CoCHPL} framework for the MTAMCR scenario by formulating MTAMCR as a hierarchical RL task, which includes an option selection task and chain-of-choice generation tasks for different options~(action types).
    \item We conduct extensive experiments on four benchmark datasets, and the results demonstrate a substantial improvement in both performance and generative capacity with all the comparison methods.
\end{enumerate}

\section{Related Work}

\subsection{Multi-round Conversational Recommendation}
Multi-round conversational recommendation (MCR) is a prevalent setting in conversational recommendation, characterized by iterative questioning and recommendations until user satisfaction or a dialogue round limit is reached~\cite{deng2021unified,lei2020interactive,xu2021adapting}. Early models like CRM~\cite{sun2018conversational} utilized reinforcement learning in single-round settings, later extended to MCR by EAR~\cite{lei2020estimation}. SCPR~\cite{lei2020interactive} considers MCR as interactive path reasoning on a graph, while UNICORN~\cite{deng2021unified} proposes a unified framework based on a dynamic weighted graph. DAHCR~\cite{zhao2023towards} introduces a hierarchical approach for action type and attribute/item selection. MCMIPL~\cite{zhang2022multiple} proposes Multiple-Choice MCR to enable multiple-choice questions within an attribute type per turn. Instead of single-type attribute queries per turn, our work introduces a more realistic setting, Multi-Type-Attribute Multi-round Conversational Recommendation.

\subsection{The Options Framework}
The options framework by Sutton et al.\cite{sutton1999between} extends the conventional reinforcement learning paradigm~\cite{zhang2022multi} by incorporating temporally extended actions, or options. Each option $\omega \in \Omega$ comprises an intra-option policy $\pi_{\omega}$ for action selection, a state transition function $\mathcal{T}_{\omega}$, and a termination function $\beta_{\omega}$ to decide when to terminate the current option. This hierarchical structure aids in efficient exploration and task decomposition. The option-critic architecture, introduced by \cite{bacon2017option}, integrates policy gradient methods with the options framework\cite{sutton1999policy}, enabling concurrent learning of intra-option policies and termination conditions, and facilitating subgoal identification without additional reward schemes.

\section{Preliminary}
\subsubsection{Multi-Type-Attribute Multi-round Conversational Recommendation.}

Generally~\cite{lei2020estimation,lei2020interactive,zhang2022multiple}, CRS engages in a dialogue with users, asking questions about attributes or providing recommendations in each round. The users are assumed to have definite preferences toward specific attributes and items, and they can express preferences by accepting or rejecting the attributes or items mentioned by CRS. The conversation continues until a successful recommendation is made or a pre-set round limit is reached. Our focus is on a more realistic CRS paradigm, Multi-Type-Attribute Multi-round Conversational Recommendation, where the CRS can question the user about multiple types of attributes per turn.

In MTAMCR, we denote the user set by $\mathcal{U}$, the item set by $\mathcal{V}$, and the attribute set by $\mathcal{P}$. Each item $v \in \mathcal{V}$ is associated with attributes $\mathcal{P}_v$, with each attribute $p \in \mathcal{P}_v$ corresponding to a specific type. In each episode, there is a target item $v$ that is acceptable to the user $u \in \mathcal{U}$. At the start of each session, a user $u \in \mathcal{U}$ targets an item $v$ and indicates a preferred attribute $p_0 \in \mathcal{P}_v$ to the CRS. Then, in each turn $T$, the CRS is free to query the user for attribute preference or recommend items from the candidates through multiple choice questions:
\begin{equation} \label{pre:choices}
    C_{T} = \{a_1, a_2, ..., a_n\},
\end{equation}
where $a$ represents a choice~(\ie attribute or item), and $C_T$ may comprise multiple attributes $\{p_1, p_2, ..., p_n\}$ or items $\{v_1, v_2, ..., v_n\}$. The user $u$ responds to each choice in $C_T$ presented by the CRS, either accepting or rejecting it.

\begin{figure}[t]
	
        \begin{minipage}{1\linewidth}
 	    \centerline{\includegraphics[width=\textwidth]{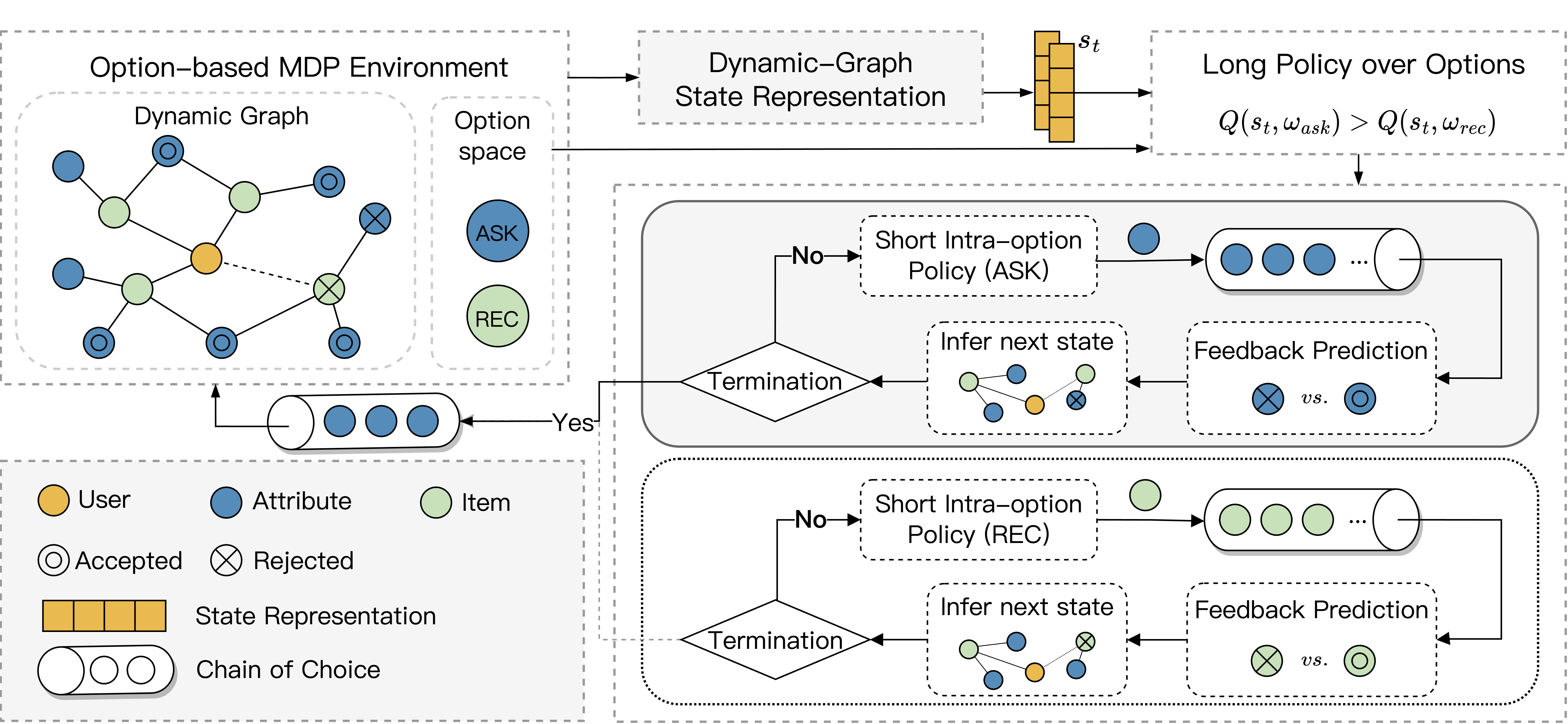}}
        \end{minipage}
 
	\caption{The overview of Chain-of-Choice Hierarchical Policy Learning.}
	\label{fig:pipeline}
\end{figure}
\section{Chain-of-Choice Hierarchical Policy Learning}
In this section, we present a detailed description of our method, CoCHPL, as illustrated in \figref{fig:pipeline}, to explain the overview of the chain-of-choice generation. CoCHPL consists of four key components: an option-based MDP Environment to provide the dynamic state with the graph structure and candidate options and choices for the agent, a Dynamic-Graph State Representation module that encodes the graph-based state into a latent space to capture the key information and relationships, a Long Policy Over Options to determine the option~(\ie ask or recommend) per round, and a Short Intra-Option Policy to generate the chain of choices via multi-step reasoning and selection.

\subsection{Option-based MDP Environment}
By formulating the MTAMCR as a hierarchical reinforcement learning task, we model not only the low-level actions~(choices) but also the high-level actions~(options) within each turn, which helps the CRS learn different dimensions of temporal abstractions. The original environment consists of a state space $S$, an action space $\mathcal{A}$, a transition function $\mathcal{T}: \mathcal{S} \times \mathcal{A} \rightarrow \mathcal{S} $ and a reward funtion $\mathcal{R}: \mathcal{S} \times \mathcal{A} \rightarrow \mathbb{R}$.
The Option-Based MDP environment extends the MDP environment to the tuple $(\mathcal{S}, \Omega, \mathcal{A}, \mathcal{T}, \mathcal{R})$ by introducing options $\Omega = \{\omega_{ask}, \omega_{rec}\}$, which are alternative action types that the agent can choose from at each turn. 

To model the state transitions between the multiple choices within the same turn, we assume that the user answers the question choices one by one and introduces the timestep $t$ to describe the intrinsic state of turn $T$ as follows:
\begin{equation}\label{method:timestep}
s_T = \{s_t, s_{t+1}, ...,s_{t+n}\}, s_{T+1} = \{s_{t+n}, s_{t+n+1}, ...,s_{t+n+m}\},
\end{equation}
where $n$, $m$ is the choice number in turn $T$ and $T+1$, respectively. If the user responds to the first choice $a_t$ at turn $T$, the state will transition from $s_t$ to $s_{t+1}$.

\subsubsection{State-Option Space}

Before each turn $T$, the CRS selects an option $\omega_T$ from $\Omega$ to decide the action type and extend the state space. The state-option pair $(s_t, \omega_T)$ at timestep $t$ in turn $T$ contains $\{u, \mathcal{V}^{(t)}, \mathcal{P}^{t}, \omega_T\}$, where $\mathcal{V}^{(t)} = \mathcal{V}^{(t)}_{rej} \cup \mathcal{V}^{(t)}_{cand}$, and $\mathcal{P}^{(t)} = \mathcal{P}^{(t)}_{acc} \cup \mathcal{P}^{(t)}_{rej} \cup \mathcal{P}^{(t)}_{cand}$. $\mathcal{P}^{(t)}_{acc}$ denotes all the accepted attributes, $\mathcal{V}^{(t)}_{rej}$ and $\mathcal{P}^{(t)}_{rej}$ denotes all the rejected items and attributes, $\mathcal{V}^{(t)}_{cand}$ and $\mathcal{P}^{(t)}_{cand}$ denotes the candidate items and attributes.

\subsubsection{Action Space.}
For $(s_t, \omega_{rec})$ at timestep $t$, the candidate action space $\mathcal{A}_t$ is $\mathcal{V}^{(t)}_{cand}$ for the option $\omega_{rec}$. We have $\mathcal{V}_{cand }^{(t)}=\mathcal{V}_{\mathcal{P}_{acc}^{(t)}} \backslash \mathcal{V}_ {rej}^{(t)}$ where $\mathcal{V}_{\mathcal{P}_{acc}^{(t)}}$ denotes all the items which contain all the accepted attributes. For another pair $(s_t, \omega_{ask})$, the candidate attribute space $\mathcal{P}_{cand}^{(t)}=\mathcal{P}_{\mathcal{V}_{cand }^{(t)}} \backslash(\mathcal{P}_{acc}^{(t)} \cup \mathcal{P}_{rej }^{(t)})$ and $\mathcal{P}_{\mathcal{V}_{cand}^{(t)}}$ denotes the attributes of all the candidate items.

\subsubsection{Transition.}

For the state-option pair $(s_t, \omega_{ask})$, after the selection of $a_t$ at timestep $t$, if the CRS predicts that the user accepts $a_t$, then $s_t$ will transiton to $s_{t+1}$ and $\mathcal{P}_{acc}^{(t+1)} = \mathcal{P}_{acc}^{(t)} \cup a_{t}$. If rejected, $\mathcal{P}_{rej}^{(t+1)}$ are updated to $\mathcal{P}_{rej}^{(t)} \cup a_{t}$. The next candidate action space is updated to $\mathcal{P}_{cand}^{(t+1)} = \mathcal{P}_{cand}^{(t)} \backslash a_{t}$.
The transition is similar for another pair $(s_t, \omega_{rec})$.

\subsubsection{Reward.}
To promote a universal and uniform distribution of rewards $\mathcal{R}$, a simplification of the reward structure is implemented as (1) $r_{ask\_suc}$: a slight reward when the user accepts the asked attribute, (2) $r_{ask\_fail}$: a slight penalty when the user rejects the asked attribute, (3) $r_{rec\_fail}$: a slight penalty when the user rejects the recommended item, (4) $r_{rec\_suc}$: a strong reward when the user accepts the recommended item.

\subsection{Dynamic-Graph State Representation Learning}
We explicitly employ a dynamic-graph state representation in the option-based MDP environment to incorporate user-preferred attributes and candidate items. This representation captures both the current conversation and its historical context, facilitating the modeling of conversational recommendation as an interactive path reasoning problem on a graph.

\subsubsection{Graph Construction.}
We denote the dynamic undirected graph as ${G}_u = (\mathbf{V}, \mathbf{A})$. The node set $\mathbf{V}$ consists of the related user, items, and attributes. And $\mathbf{A}$ is a $n \times n$ adjacency matrix representing each node's edge weight in $\mathbf{V}$. During each timestep in the conversation, $G_u$ dynamically changes the node and adjacency matrix as follows:

\begin{equation}\label{method:node}
    \mathbf{V}^{(t)}=\{u\} \cup \mathcal{P}_{acc}^{(t)} \cup  \mathcal{P}_{cand}^{(t)} \cup \mathcal{P}_{{rej }}^{(t)} \cup \mathcal{V}_{cand}^{(t)} \cup \mathcal{V}_{ {rej}}^{(t)},
\end{equation}

\begin{equation}\label{method:edge}
\mathbf{A}_{i, j}^{(t)}=\left\{\begin{array}{ll}
w_{v}^{(t)}, & \text { if } n_{i}=u, n_{j} \in \mathcal{V}_{{cand}}^{(t)},\\
1, & \text { if } n_{i} \in \mathcal{V}_{{cand}}^{(t)}, n_{j} \in \mathcal{P},\\
0, & \text { otherwise },
\end{array}\right.
\end{equation}
where $w_{v}^{(t)}$ denotes the user's preference score for the item $v$, calculated as:
\begin{equation}\label{method:preferscore}
w_{v}^{(t)}=\sigma(e_{u}^{\top} e_{v} + \sum\nolimits_{p \in \mathcal{P}_{acc}^{(t)}} e_{v}^{\top} e_{p} - \sum\nolimits_{p \in \mathcal{P}_{rej}^{(t)}} e_{v}^{\top} e_{p}),
\end{equation}
where $\sigma(\cdot)$ is the sigmoid function and ${e_{u}, e_{v},e_{p}}$ are pre-trained embeddings initialized as~\cite{bordes2013translating}. We can also calculate the preference score for candidate attributes at the current state: $w_{p}^{(t)}=\sigma(e_{u}^{\top} e_{p} + \sum\nolimits_{p \in \mathcal{P}_{acc}^{(t)}} \overline{e}_{p}^{\top} e_{p} - \sum\nolimits_{p \in \mathcal{P}_{rej}^{(t)}} \overline{e}_{p}^{\top} e_{p})$. With the calculated weights, we can reduce the action space size for different options by selecting the top-$K$ candidates in the action selection section~\ref{section:longpolicy}, \ref{section:shortpolicy}.

\subsubsection{State Representation.}
By incorporating both the current conversation and the historical conversation data, we can create a comprehensive global knowledge graph denoted as $G$. This knowledge graph captures the evolving preferences of users over time. To effectively leverage the correlation information among users, items, and attributes represented in the graph, we implement a two-layer graph convolutional network (GCN)~\cite{kipf2016semi,zhang2020semi} as follows:
\begin{equation} \label{method:gcn}
    e_{s}^{(l+1)}=\operatorname{ReLU}(\sum\nolimits_{i \in \mathcal{N}_{s}} D^{-\frac{1}{2}} \boldsymbol{A} D^{-\frac{1}{2}}  W^{l} e_{i}^{(l)} + B^{(l)}e_{s}^{(l)}),
\end{equation}
where $l$ is the layer number, $\mathcal{N}_{s}$ denotes the neighbors of $n_s$ and $D$ denotes the degree matrix with $D_{ii} = \sum\nolimits_j \boldsymbol{A}_{i,j}$. $B^{(l)}$ and $W^{(l)}$ are both trainable network parameters. To further extract the sequential information of the current conversation, we use a single-layer multi-head attention~\cite{vaswani2017attention} network and calculate the final state representation by an average pooling:
\begin{equation} 
\mathcal{P}_{acc}^{*} = \operatorname{LayerNorm}(\operatorname{FNN}(\operatorname{MultiHead}(\mathcal{P}_{acc}^{(t)})+\mathcal{P}_{acc}^{(t)})) \label{method:multihead}
\vspace{-1mm}
\end{equation}
\begin{equation}
f_{\theta_{S}}(s_t) = \operatorname{AvgPool}(\mathcal{P}_{acc}^{*}). \label{method:avgpool}
\end{equation}
After the sequential learning, the final state representation is generated, and $\theta_{S}$ denotes all parameters for the dynamic-graph state representation module.

\begin{algorithm}[!ht]
	\caption{{CoCHPL in the MTAMCR scenario}}
	\label{algo1}
            Parameter initialization: $\theta_{Q}$, $\theta_{\beta}$, $\theta_{\mathcal{T}}$, $\theta_{S}$\\
            \For{$episode=1$ \KwTo $K$}
            {
                Initialization: User $u\in\mathcal{U}$, Target item $v\in\mathcal{V}_u$, Initial attribute $p_0 \in \mathcal{P}_v$\\
                Initialization: Timestep $t \leftarrow 0$, Turn number $T \leftarrow 0$\\
                Initialization: Choice number $n \leftarrow 0$, Initial state $s_{t+n} \leftarrow s_0$\\
                Choose an action option $\omega_T$ via $\epsilon$-soft long policy $\pi_{\Omega}(s_{t+n})$\\
		      \Repeat{$\text{recommended successfully}$ or $T > T_{max}$}{
                    \text{}\\
                    \textbf{1. Chain-of-Choice Generation:}\\
                        Initialization: choice chain ${C}_T \leftarrow \{\}$, candidate action space $\mathcal{A}_{t}$\\
                        
                        \Repeat{$\beta_{\omega_T}(s_{t+n})$ terminates or  ${C}_T$ reach the maximum length}{
                        Choose an choice $a_{t+n}$ via short intra-option policy $\pi_{\omega_T}(s_{t+n})$\\
                        \uIf{at training stage}{
                            Take choice $a_{t+n}$, receive true reward $r_{t+n}$\\
                            Observe true $s_{t+n+1}$, get candidate action space $\mathcal{A}_{t+n+1}$\\
                            Store $(s_{t+n}, a_{t+n}, r_{t+n}, s_{t+n+1}, \mathcal{A}_{t+n+1}, \omega_T)$
                        }
                        \Else{
                            Predict user feedback and reward via $\mathcal{T}_{\omega}(s_{t+n}, a_{t+n})$\\
                            Update the next state $s_{t+n+1}$ based on predicted feedback\\
                        }
                        Update $C_T \leftarrow C_T+a_{t+n}$, $\, n \leftarrow n+1$
                        }
                    \text{}\\
                    \textbf{2. User Interaction:}\\
                    Take chain of choice $C_T$, observe $s_{t+n}$\\
                    Choose new option $\omega_{T+1}$ via $\epsilon$-soft $\pi_{\Omega}(s_{t+n})$\\
                    Update $t \leftarrow t+n$, $n \leftarrow 0$, $T \leftarrow T+1$

                    \text{}\\
                    \textbf{3. Policy Optimization:}\\
                    Sample mini-batch of $(s_{t+n}, a_{t+n}, r_{t+n}, s_{t+n+1}, {A}_{t+n+1}, \omega_T)$\\
                    Update $\theta_{Q}, \theta_{S}$ \wrt \equref{loss:Qvalue}\\
                    Update $\theta_{\beta}$ \wrt \equref{loss:termination}\\
                    Update $\theta_{\mathcal{T}}$ \wrt \equref{loss:feedback}\\
                }
            }
\end{algorithm}

\subsection{Long Policy Leaning Over Options}\label{section:longpolicy}
After encoding the state with the dynamic graph, we propose a long-term policy denoted as $\pi_{\Omega}$ to select the option $\omega$ at each turn. Our objective is to optimize the discounted return at turn $T$ directly. To achieve this, we express the expected Q-value of each option as follows:
\begin{equation}\label{method:longvalue}
Q_{\Omega}(s_t, \omega_T)=\sum\nolimits_{a_t \in \mathcal{A}_t} \pi_{\omega_T}(a_t \mid s_t) Q_U(s_t, \omega_T, a_t),
\end{equation}
where $\pi_{\omega_T}$ is the intra-option policy and $Q_U: \mathcal{S} \times \Omega \times \mathcal{A} \rightarrow \mathbb{R}$ is the estimated Q-value of executing an action in the context of a state-option pair which is introduced in section~\ref{section:shortpolicy}.

\subsection{Short Intra-Option Policy Learning}\label{section:shortpolicy}
With the state-option pair, we need a short-term policy to generate the choice chain at each turn.
We propose that each option $\omega$ consists of a triple $(\pi_\omega, \beta_\omega, \mathcal{T}_\omega)$ in which $\pi_\omega \in \{\pi_{\omega_{ask}},\pi_{\omega_{rec}}\}$ is a short-term intra-option policy to select the attribute or item at each timestep, $\mathcal{T}_\omega: \mathcal{S} \times \mathcal{A} \rightarrow [0, 1]$ is a feedback prediction function to predict user response for inferring the next timestep state, and $\beta_{\omega}: \mathcal{S} \rightarrow [0,1]$ is a termination function to determine when to exit the generation process of the current option.

As shown in \algoref{algo1}, during each timestep $t$ in turn $T$, $\pi_\omega$ select a item or attribute as the choice $a_t$, and then predict the feedback~(\ie acc/rej) via $\mathcal{T}_{\omega}(s_t, a_t) = \mathds{1}(a_t \text{ is accpeted})$ for inferring the next state $s_{t+1}$. If $\beta_{\omega_T}(s_{t+1})$ determines to terminate the choice chain generation, the CRS will present the chain-of-choice question to the user. Otherwise, the generation will continue.

To optimize intra-option policy $\pi_\omega$, we estimate the Q-value of selecting a choice in the context of a state-option pair $(s, \omega)$, which can be calculated by:
\begin{equation} \label{method:shortvalue}
	Q_U(s_t, \omega_T, a_t) = r(s_t, a_t)+\gamma U(\omega_{T}, s_{t+1}),
\end{equation}
where $r: \mathcal{S}\times \mathcal{A} \rightarrow \mathbb{R}$ is the reward function and $U: \Omega \times \mathcal{S} \rightarrow \mathbb{R}$ denotes the value function of executing option $\omega_{T}$ upon entering the next state $s_{t+1}$ as:
\begin{equation}\label{method:Uvaluefunction}
    U(\omega_{T}, s_{t+1}) =(1-\beta_{\omega_T}(s_{t+1})) Q_\Omega(s_{t+1}, \omega_T)+\beta_{\omega_T}(s_{t+1})\max \limits_{\overline{\omega}}Q_\Omega(s_{t+1}, \overline{\omega}),
\end{equation}
where the next state value is estimated by a weighted sum of the value of continuing the current option and the maximum value in the next state, with the weights determined by the termination value.

\subsection{Model Training}
As illustrated in \algoref{algo1}, in MTAMCR, the CRS needs to generate a complete chain of choices before interacting with the user and obtaining feedback. However, during the training stage, we can receive real feedback for each choice $a_t$ and reward $r_t$ immediately from users at each timestep $t$, and accurately infer the next state $s_{t+1}$, which speeds up the convergence of training. Each option $\omega$ has its own experience replay buffer, denoted by $\mathcal{D}{\omega}$, and the experience $d_t = (s_t, a_t, r_t, s_{t+1}, \mathcal{A}_{t+1}, \omega_T)$ is stored in the replay buffer $\mathcal{D}{\omega_T}$. We sample a mini-batch of experiences from each $\mathcal{D}_{\omega_T}$ after every turn and optimize the parameters of the whole network.

In this work, we implement the long policy over options $\pi_\Omega$ and short intra-option policies $\pi_\omega$ to sample the probability distribution of the discrete action space with the Q-value function as follows:

\begin{equation}\label{method:longshortpolicy}
\pi_{\Omega}(s_t) = \mathop{\arg\max}_{\omega} Q_{\Omega}(s_t, \omega), \pi_{\omega_T}(a_t \mid s_t) = \frac{\exp (Q_{U}(s_t, \omega_T, a_t))}{\sum_{a \in \mathcal{A}_t} \exp (Q_{U}(s_t, \omega_T, a))},
\end{equation}
where both $\pi_{\Omega}$ and $\pi_{\omega}$ freeze the parameters for the stable training optimization and only need a learnable $Q_{U}$ function to represent $\pi_{\Omega}$, $\pi_{\omega}$ and $Q_{\Omega}$.We devise a dueling deep Q-network~\cite{wang2016dueling} to estimate the expected value of the triple $(s_t, \omega_t, a_t)$ at timestep $t$ and the Q-value of the tuple can be estimated by:
\begin{equation} \label{method:Qshortoption}
	\begin{split}
	Q_{U}(s_t, \omega_T, a_t;\theta_Q,\theta_S)= f_{\theta_{V}}(f_{\theta_{S}}(s_t))+f_{\theta_{A}}(f_{\theta_{S}}(s_t), \omega_t, a_t),
	\end{split}
\end{equation}
where $\theta_{V}$ and $\theta_{A}$ represent the parameters of the value and advantage functions, respectively, simplified as $\theta_{Q} = \{\theta_{V}, \theta_{A}\}$. The target value $y_{t}$ of $Q_{U}$ during training can be obtained as \equref{method:shortvalue}, and we can optimize the parameters of the Q network by minimizing the MSE loss as follows:
\begin{equation} \label{loss:Qvalue}
	\mathcal{L}(\theta_Q,\theta_S)=\mathbb{E}_{d_t \sim \mathcal{D}}[(y_{t}-Q_{U}(s_t, \omega_T, a_t;\theta_Q,\theta_S))^{2}].
\end{equation}

Based on the Termination Gradient Theorem proposed by~\cite{bacon2017option}, the parameters $\theta_\beta$ of termination network $\beta_\omega$ can be optimized by increasing the gradient or minimizing the loss:

\begin{equation} \label{loss:termination}
	\mathcal{L}(\theta_\beta)=\mathbb{E}_{d_t \sim \mathcal{D}}[\beta_\omega(s_{t+1};\theta_\beta)(Q_\Omega(s_{t+1},\omega_T) - V(s_{t+1}))],
\end{equation}
where $V(s_{t+1}) = f_{\theta_{V}}(f_{\theta_S}(s_{t+1})$ is the estimated value of the next state. Intuitively, when the value of selecting the current option exceeds the estimated value, the probability of terminating the option is reduced.

For the feedback prediction function, we simply implement it by a multilayer perceptron~(MLP). With the ground-truth state and reward, we optimize it by minimizing the following loss to minimize the squared difference between the predicted feedback for action $a_t$ and the ground-truth user feedback:
\begin{equation} \label{loss:feedback}
    \mathcal{L}(\theta_\mathcal{T})=\mathbb{E}_{d_t \sim \mathcal{D}}[(\mathds{1}(a_t \text{ is accepted})-\mathcal{T}_{\omega_t}(s_t, a_t;\theta_\mathcal{T}))^{2}],
\end{equation}
where $\mathds{1}(\cdot)$ takes the value of 1 when user accepts $a_t$, and 0 otherwise.
\setcounter{footnote}{0}
\section{Experiments}
To demonstrate the superiority of CoCHPL
in the MTAMCR scenario, we conduct experiments aimed at validating the following three research questions~(RQs):
\begin{itemize}
  \item \textbf{RQ1:} With the comparison of state-of-the-art methods, how much does our model CoCHPL improve in overall performance?
  \item \textbf{RQ2:} How do different components contribute to the performance?
  \item \textbf{RQ3:} Does CoCHPL enhance attribute diversity and dependency within each conversational turn?
\end{itemize}

\subsection{Experimental Settings}
\subsubsection{Datasets.}
To fairly evaluate CoCHPL, we utilized four widely recognized benchmark datasets~\cite{zhang2022multiple} that are commonly used for conversational recommendation tasks involving multiple choice questions, namely:
LastFM, Yelp, Amazon-Book and MovieLens.
The statistics of datasets are shown in Table~\ref{tab:datasets}.
\begin{itemize}
    \item \textbf{LastFM} is a popular online platform that focuses on music recommendations and social networking.
The original attributes in LastFM were treated as attribute instances, and a clustering method was utilized to identify 34 attribute types.
    \item \textbf{Yelp} serves for business recommendations. In the context of Yelp, a two-layer taxonomy was created by \cite{lei2020estimation}, comprising 29 attribute types and 590 attribute instances. 
The attribute types in the taxonomy represent different categories or aspects of businesses, such as price range, ambiance, service quality, and more. 
    \item \textbf{Amazon-Book} is widely used for book recommendations. Users and items with a minimum of 10 interaction records are retained. Entities and relations in the knowledge graph are considered as attribute instances and attribute types, respectively.
    \item \textbf{MovieLens} is a rating dataset that contains user ratings for movies. Interactions with ratings above three are retained, and the knowledge graph entities and relations are considered as attribute instances and attribute types.
\end{itemize}

\begin{table}[t]
\centering
\renewcommand{\arraystretch}{1.2}
\setlength{\tabcolsep}{2mm}
\caption{Statistics of four benchmark datasets.}
\label{tab:datasets}
\begin{tabular}{@{}lrrrr@{}}
\toprule
               & \textbf{LastFM} & \textbf{Yelp} & \textbf{Amazon-Book} & \textbf{MovieLens}  \\ \midrule
\#Users        & 1,801           & 27,675        & 30,291      &   20,892                          \\
\#Items        & 7,432           & 70,311        & 17,739      &    16,482                        \\
\#Attributes   & 8,438           & 590           & 988         &    1,498                        \\
\#Types        & 34              & 29            & 40          &    24                             \\ 
\midrule
\#Entities     & 17,671          & 98,576        & 49,018       &   38,872                      \\
\#Relations    & 4               & 3             & 2            &   2                      \\
\#Interactions & 76,693          & 1,368,606     & 478,099     &   454,011                        \\
\#Triples     & 228,217         & 2,533,827     & 565,068      &  380,016 \\
\toprule
\end{tabular}
\end{table}

\subsubsection{Baselines.}
To benchmark our model's performance, we compared it against eight established baselines:
\begin{itemize}
    \item \textbf{Abs Greedy}~\cite{christakopoulou2016towards} only recommends items in each turn until the user accepts the recommendation or the total turn number exceeds the maximum limit.
    \item \textbf{Max Entropy}~\cite{lei2020estimation} asks or recommends the top-scored instances with the maximum entropy with a certain probability.
    \item \textbf{CRM}~\cite{sun2018conversational} introduces a policy that selects the action and terminates the conversation when the recommendation is rejected.
    \item \textbf{EAR}~\cite{lei2020estimation} proposes a three-stage pipeline called Estimation–Action-Reflection for selecting actions and optimizing the model.
    \item \textbf{SCPR}~\cite{lei2020interactive} employs graph-based path reasoning to retrieve actions and graph embedding to represent user preferences dynamically.
    \item \textbf{UNICORN}~\cite{deng2021unified} offers a comprehensive approach for selecting optimal actions, integrating both the conversation and recommendation components.
    \item \textbf{MCMIPL}~\cite{zhang2022multiple} focuses on single-type attribute selection in multiple-choice question settings and uses a multi-interest extractor for preference capture.
    \item \textbf{DAHCR}~\cite{zhao2023towards} employ a director to select the action type and an actor to choose the asked attribute or recommended item.
\end{itemize}

To ensure a fair comparison in the MTAMCR scenario, we made two kinds of modifications to UNICORN, MCMIPL, and DAHCR:
\newpage
\begin{itemize}
    \item \textbf{Single-Type attributes (-S)}: Also known as attribue type-based multiple choice~\cite{zhang2022multiple}. In each round, CRS selected one attribute type and then chose multiple attribute instances within that type.
    \item \textbf{Multi-Type attributes (-M)}: In each round, we selected the top-K attribute instances with the highest Q-values, which could be from different or the same attribute types.
\end{itemize}

\begin{table}[t]
\centering
\renewcommand{\arraystretch}{1.2}%
\caption{Overall performance evaluated by \texttt{SR@15}, \texttt{AT}, and \texttt{hDCG} across four datasets.}
\label{tab:overall_performance}
\resizebox{\textwidth}{!}{
\begin{tabular}{lcccccccccccc} 
\toprule[1pt]
\specialrule{0em}{0.5pt}{0.5pt}
               \multirow{2}{*}{\textbf{Model}} & \multicolumn{3}{c}{\textbf{LastFM}}                      & \multicolumn{3}{c}{\textbf{Yelp}}                         & \multicolumn{3}{c}{\textbf{Amazon-Book}}                 & \multicolumn{3}{c}{\textbf{MovieLens}}                    \\
\specialrule{0em}{0.5pt}{0.5pt}
\cmidrule(r){2-4}\cmidrule(r){5-7}\cmidrule(r){8-10}\cmidrule(r){11-13}\\
\specialrule{0em}{-8pt}{-6pt}
               & SR@15          & AT            & hDCG           & SR@15          & AT             & hDCG           & SR@15          & AT            & hDCG           & SR@15          & AT            & hDCG            
               \\
\specialrule{0em}{0pt}{0pt}
\hline
Abs Greedy\cite{christakopoulou2016towards}     & 0.635          & 8.66          & 0.267          & 0.189          & 13.43          & 0.089          & 0.193          & 13.90         & 0.087          & 0.724          & 5.24          & 0.465           \\
Max Entropy\cite{lei2020estimation}    & 0.669          & 9.33          & 0.269          & 0.398          & 13.42          & 0.121          & 0.382          & 12.43         & 0.149          & 0.655          & 6.87          & 0.412           \\
CRM\cite{sun2018conversational}            & 0.580          & 10.79         & 0.224          & 0.177          & 13.69          & 0.070          & 0.354          & 12.81         & 0.127          & 0.626          & 7.43          & 0.433           \\
EAR\cite{lei2020estimation}            & 0.595          & 10.51         & 0.230          & 0.182          & 13.63          & 0.079          & 0.411          & 11.62         & 0.153          & 0.749          & 6.28          & 0.459           \\
SCPR\cite{lei2020interactive}           & 0.709          & 8.43          & 0.317          & 0.489          & 12.62          & 0.159          & 0.432          & 11.03         & 0.148          & 0.823          & 4.94          & 0.514           \\
UNICORN\cite{deng2021unified}        & 0.788          & 7.58          & 0.349          & 0.520          & 11.31          & 0.203          & 0.498          & 10.42         & 0.179          & 0.863          & 4.62          & 0.523           \\ 
DAHCR\cite{zhao2023towards}       & 0.925          & 6.31          & 0.431          & 0.626          & 11.02          & 0.192          & 0.522          & 10.14          & 0.188          & 0.875          & 4.34          & 0.535           \\
\hline
UNICORN-S    & 0.884          & 6.07          & 0.403          & 0.593          & 10.57          & 0.212          & 0.538          &  10.01         &  0.253        &  0.883         &  4.05        &   0.585        \\
MCMIPL-S     & 0.942          & 5.44          & 0.431          & 0.631          & 10.25          & 0.225          & 0.556          & 9.57          & 0.264          & 0.879          & 3.92          & 0.581           \\
DAHCR-S       & 0.933          & 5.67          & 0.429          & 0.668          & 10.15          & 0.232          & 0.572          & 9.45          &  0.272       &  0.895         &  3.89         &  0.603          \\
\hline
UNICORN-M    & 0.892          &  5.96         & 0.426          & 0.634          & 10.35          & 0.233          & 0.584          & 9.69          &  0.261         &  0.887         & 4.10         &  0.587         \\
MCMIPL-M     & 0.973          & 4.59          & 0.466          & 0.675         & 10.11          & 0.256          & 0.667          & 8.98          & 0.293         & 0.891          & 3.68          & 0.609           \\
DAHCR-M       & 0.965          & 4.61          & 0.451         & 0.681          & 10.23          & 0.241          & 0.689          & 8.45         & 0.301          & 0.874         & 3.98          & 0.593           \\
\hline
\textbf{CoCHPL} & \underline{\textbf{0.991}} & \underline{\textbf{3.75}} & \underline{\textbf{0.493}} & \underline{\textbf{0.846}} & \underline{\textbf{9.36}} & \underline{\textbf{0.267}} & \underline{\textbf{0.775}} & \underline{\textbf{7.61}} & \underline{\textbf{0.330}} & \underline{\textbf{0.908}} & \underline{\textbf{2.85}} & \underline{\textbf{0.628}}  \\
\specialrule{0em}{0pt}{0pt}
\bottomrule[1pt]
\end{tabular}
}
\end{table}

\subsubsection{Evaluation Metrics.}
We evaluate the performance of our MCR recommendation system using three key metrics: (1) Success rate (\textbf{SR@t}) measures the cumulative ratio of successful recommendations within a maximum of $t$ turns. (2) Average turn (\textbf{AT}) evaluates the average number of terminated conversation turns for all sessions. Lower values indicate higher efficiency in the recommendation process. (3) \textbf{hDCG@(T, K)}~\cite{deng2021unified} assesses the ranking performance of recommended items by considering the target item's rank in the recommendation list, which extends the normalized discounted cumulative gain metric.

\subsubsection{Training Details.}
We set the maximum turns $T$ to 15, and the agent can recommend up to $k_v=10$ items or ask $k_p=3$ attribute-related questions at every turn. We pre-train the graph nodes with TransE \cite{bordes2013translating}, and train each dataset for $10,000$ episodes online. Adam optimizer with the learning rate $1e-4$ and discount factor $\gamma$ is set to be 0.999. For our method CoCHPL, the reward settings were as follows: $r_{ask\_acc}=1e-2$, $r_{ask\_rej}=-1e-4$, $r_{rec\_rej}=-1e-4$, and $r_{rec\_suc}=1$. We found that the effect of retrieving accepted attributes was minimal in MovieLens, and then the reward for $r_{ask\_acc}$ was adjusted to $1e-5$ for it. The rewards for all state-of-the-art baselines remain unchanged as~\cite{zhang2022multiple}.

\subsection{Experimental Results}
\subsubsection{Overall Performance (RQ1).}

\begin{figure}[t]
    \begin{minipage}{0.5\linewidth}
        \centering
        \includegraphics[width=0.9\textwidth]{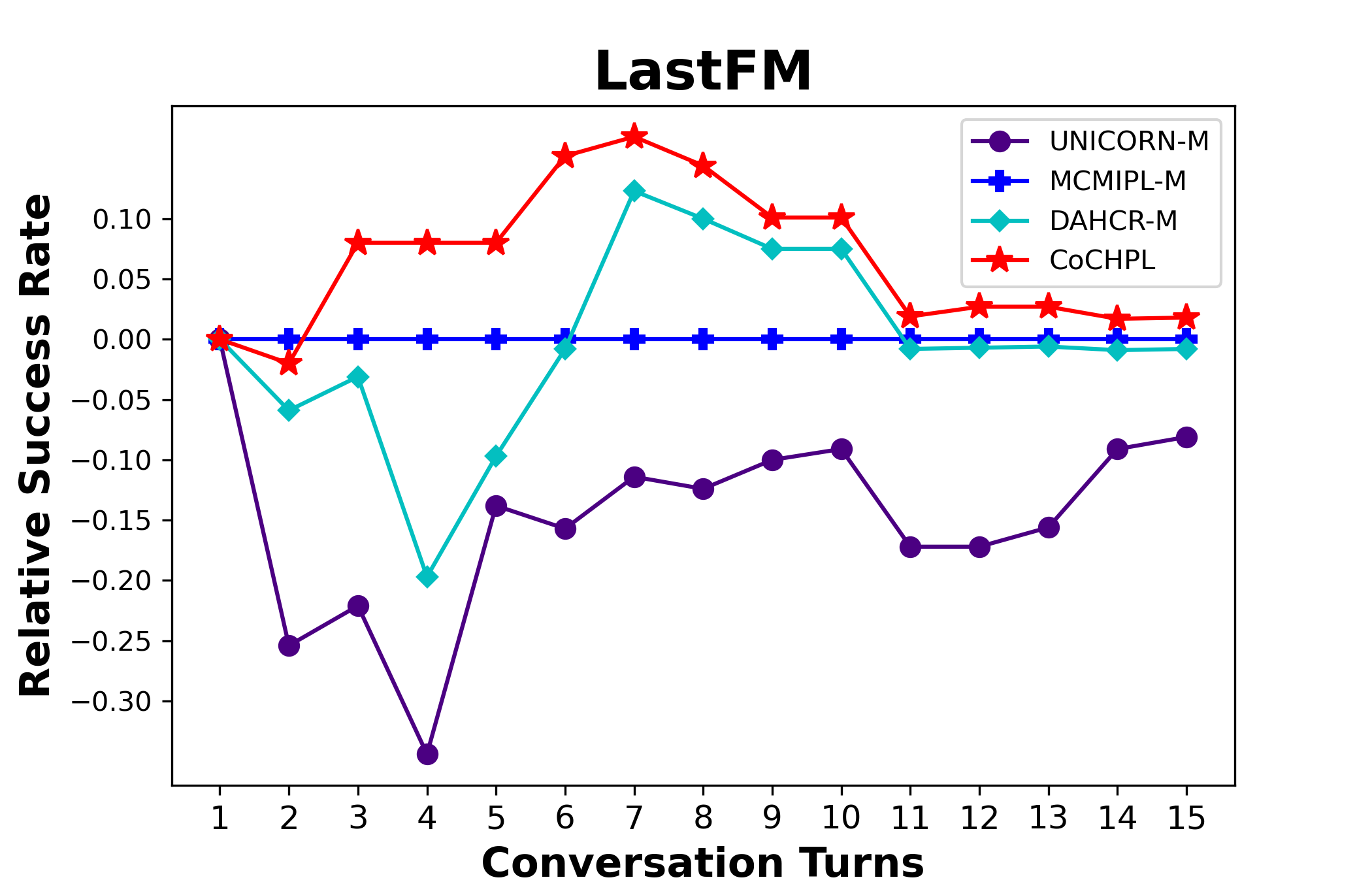}
    \end{minipage}
    \begin{minipage}{0.5\linewidth}
        \centering
        \includegraphics[width=0.9\textwidth]{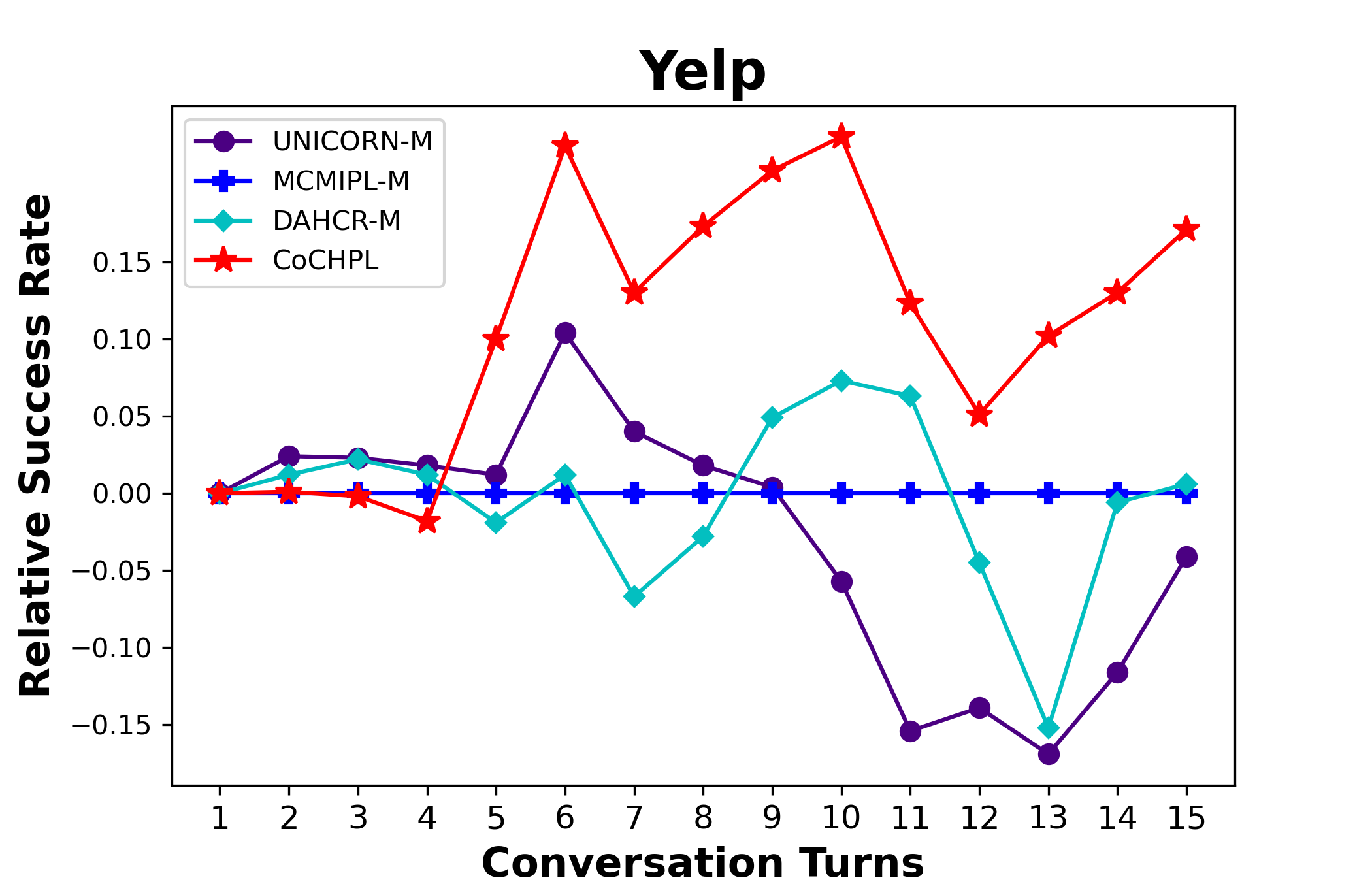}
    \end{minipage}
    
    \begin{minipage}{0.5\linewidth}
        \centering
        \includegraphics[width=0.9\textwidth]{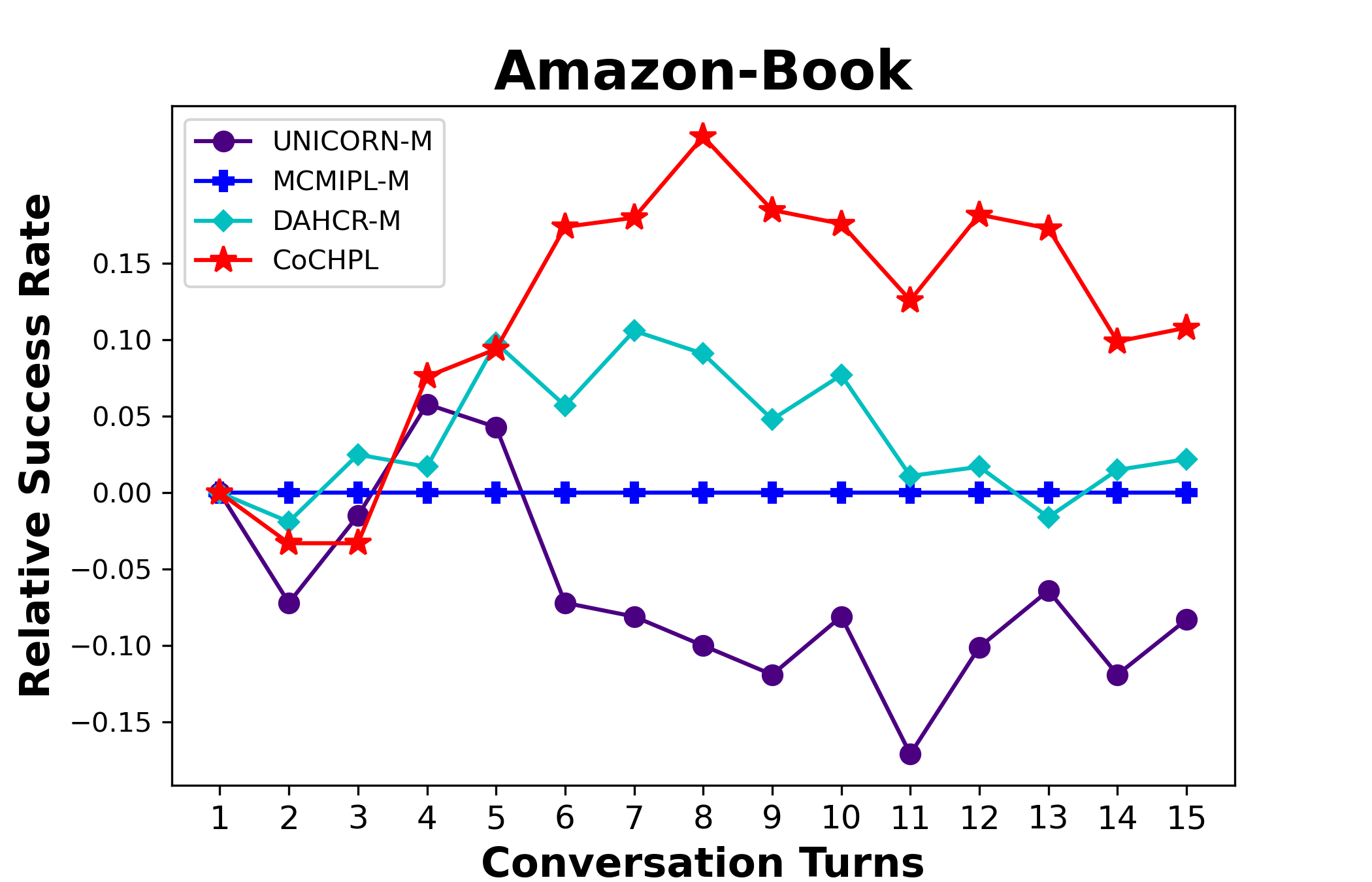}
    \end{minipage}
    \begin{minipage}{0.5\linewidth}
        \centering
        \includegraphics[width=0.9\textwidth]{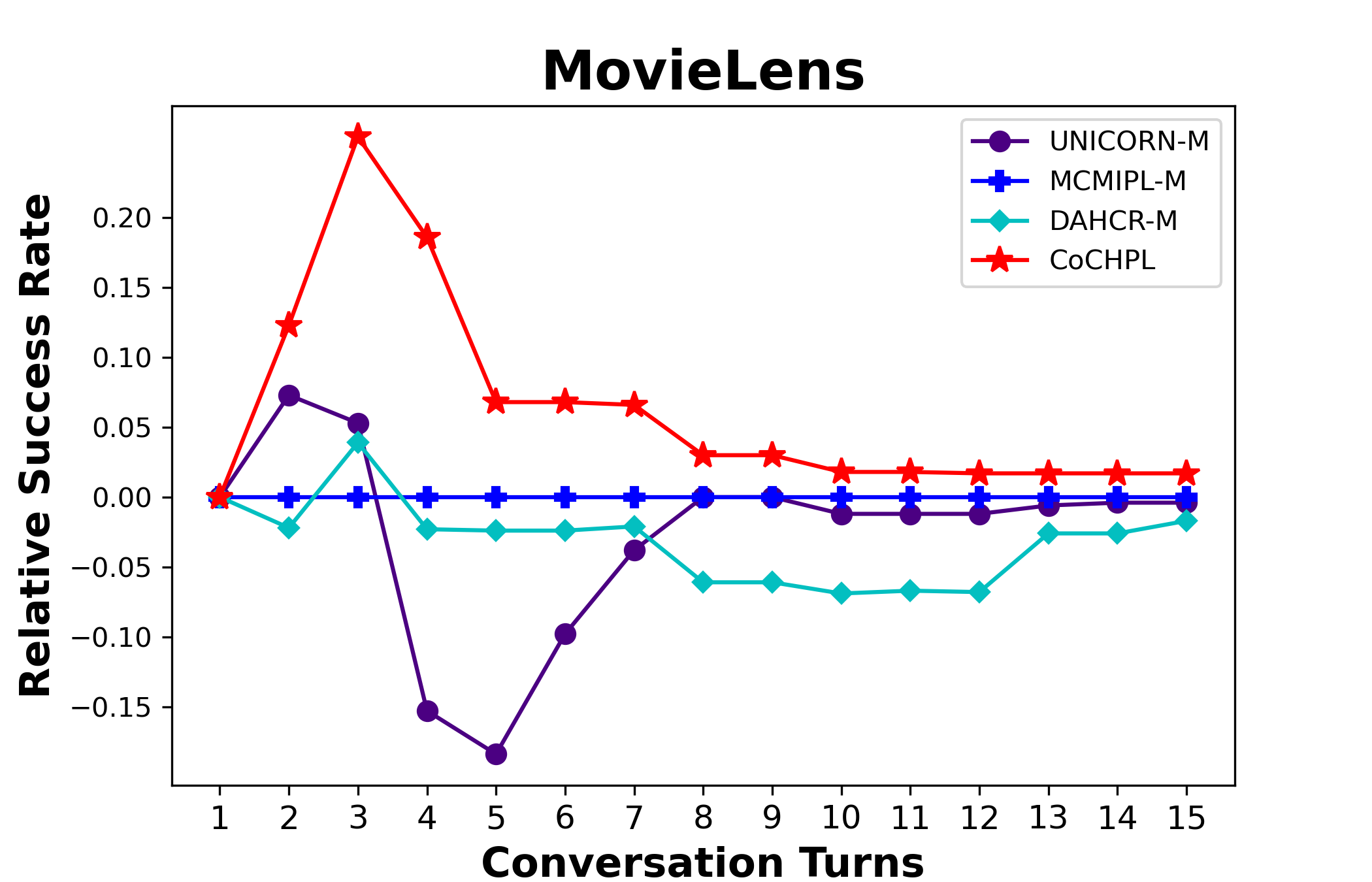}
    \end{minipage}
    
    \caption{Comparisons at different conversation turns on four datasets.}
    \label{fig:relative_success_rate}
\end{figure}

As detailed in \tabref{tab:overall_performance}, CoCHPL surpasses eight baselines across SR@15, AT, and hDCG metrics. Baselines with binary yes/no question settings exhibit lower efficiency and longer turn requirements for successful recommendations (\figref{fig:intro}(a)). In single-type attributes scenarios, UNICORN-S, MCMIPL-S, and DAHCR-S show modest improvements, while in multi-type attributes contexts, UNICORN-M, MCMIPL-M, and DAHCR-M display enhanced success rates and reduced average turns, which indicates the effectiveness of our MTAMCR approach in preference capture. 
CoCHPL's chain-of-choice generation leads to marked improvements in success rates and efficiency across benchmark datasets, notably reducing the number of turns for successful recommendations. In critical datasets like Yelp and Amazon-Book, CoCHPL demonstrates significant advancements over top methods with relative increases of $18.3\%$, $8.6\%$, and $6.9\%$ in SR@15, AT, and hDCG, respectively. For a clear comparison, we benchmarked against MCMIPL-M's SR, with differences in SR among methods serving as a measure of relative success rate (\figref{fig:relative_success_rate}) and the comparison highlights the superior performance of CoCHPL, especially in Yelp and Amazon-Book, where extensive attribute preference extraction is essential.

\begin{table}[t]
\centering
\renewcommand{\arraystretch}{1.2}%
\setlength{\tabcolsep}{2mm}

\caption{Ablation Study evaluated across Yelp and Amazon-Book datasets.}
\label{tab:ablation_study}
\resizebox{\textwidth}{!}{
\begin{tabular}{lcccccc} 
\toprule[1pt]
\specialrule{0em}{0.5pt}{0.5pt}
               &\multicolumn{3}{c}{\textbf{Yelp}}                         & \multicolumn{3}{c}{\textbf{Amazon-Book}}\\
\specialrule{0em}{0.5pt}{0.5pt}
\cmidrule(r){2-4}\cmidrule(r){5-7}\\
\specialrule{0em}{-8pt}{-6pt}
& SR@15          & AT            & hDCG           & SR@15          & AT            & hDCG            \\
\specialrule{0em}{0pt}{0pt}
\hline
\textbf{CoCHPL}  & \underline{\textbf{0.846}} & \underline{\textbf{9.36}} & \underline{\textbf{0.267}} & \underline{\textbf{0.775}} & \underline{\textbf{7.61}} & \underline{\textbf{0.330}} \\
\hline
w/o long policy over options          & 0.693          & 10.21          & 0.225         & 0.649          &   9.12       &   0.278                 \\
w/o short intra-option policy~(ask)           & 0.655           & 10.29         & 0.221         & 0.612          &  9.54         &  0.270                    \\
w/o short intra-option policy~(rec)       & 0.756          & 10.05          & 0.245          &  0.685         & 8.99          & 0.285                \\
w/o termination function                  & 0.798          & 9.67          & 0.253          & 0.744          & 8.34          & 0.302                     \\
w/o feedback prediction function          & 0.782          & 9.95          & 0.248          & 0.732          & 8.12          & 0.311             \\
\specialrule{0em}{0pt}{0pt}
\bottomrule[1pt]
\end{tabular}
}
\end{table}
\subsubsection{Ablation Study (RQ2).}
We performed an ablation study on the Yelp and Amazon-Book datasets to assess the impact of different components on system performance as follows:
(1) \texttt{w/o policy over options}: The CRS randomly selects options (action types) in each turn.
(2) \texttt{w/o intra-option policy (ask)}: The CRS chooses the top-$K$ attribute instances with the highest Q-value at every turn.
(3) \texttt{w/o intra-option policy (recommend)}: The CRS selected the top-$K$ items with the highest Q-value at each turn.
(4) \texttt{w/o termination function}: The CRS asks for or recommends the maximum number of instances in each turn without appropriate termination.
(5) \texttt{w/o feedback prediction function}: The CRS randomly predicts whether the user would accept or reject the previous choice.
From \tabref{tab:ablation_study}, we can observe that \texttt{short intra-option policy (ask)} and \texttt{long policy over options} play crucial roles in performance improvement, which suggests that learning how to select options and effectively generate multiple-choice questions about attributes is essential in the MTAMCR scenario.
Additionally, the significance of the termination and feedback prediction functions underscores the importance of learning appropriate termination conditions and the necessity of simulating user behavior during the chain-of-choice generation process.

\begin{figure}[t]
	\begin{minipage}{0.5\linewidth}
		\centerline{\includegraphics[width=0.9\textwidth]{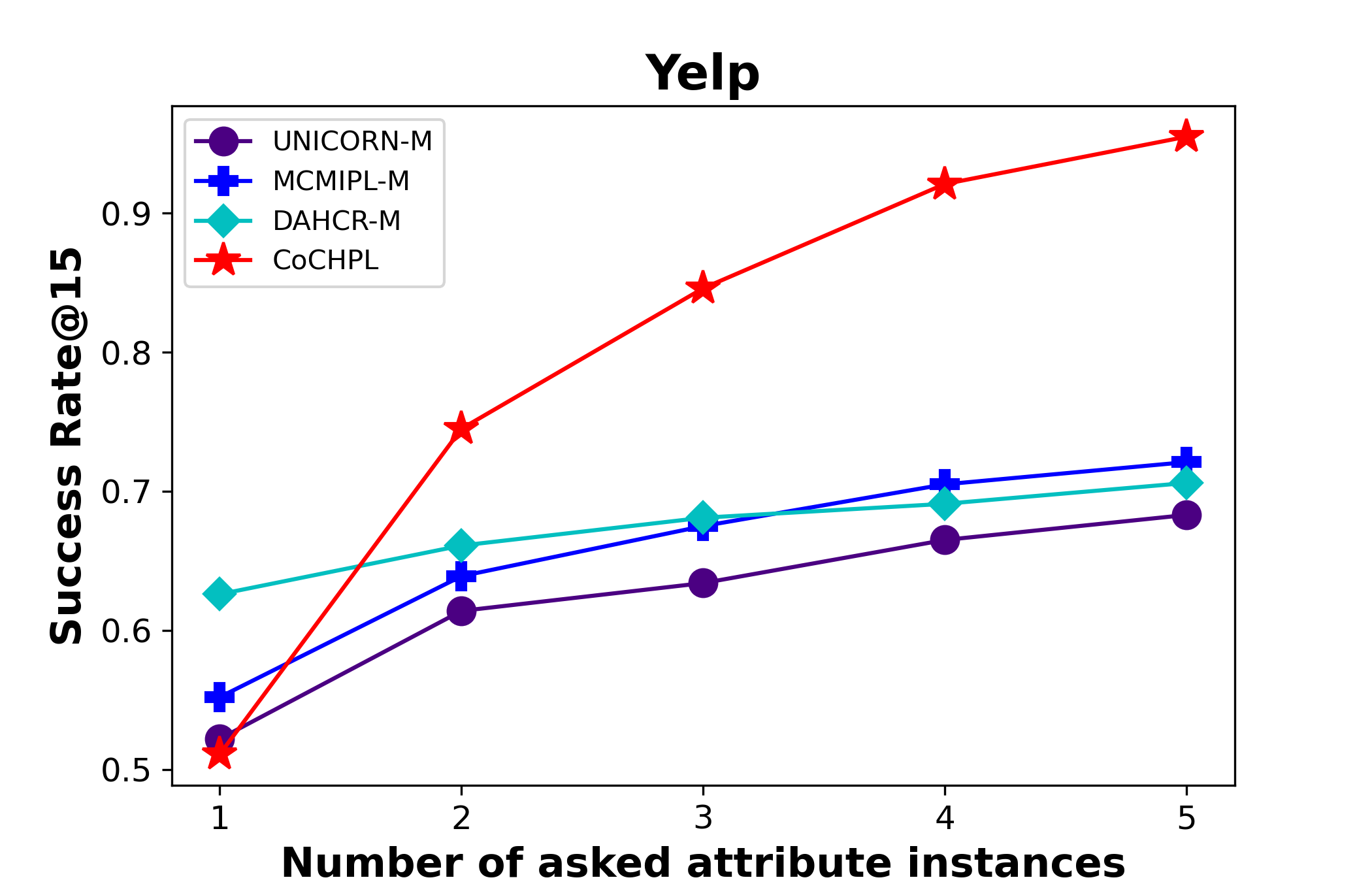}}
	\end{minipage}
	\begin{minipage}{0.5\linewidth}
		\centerline{\includegraphics[width=0.9\textwidth]{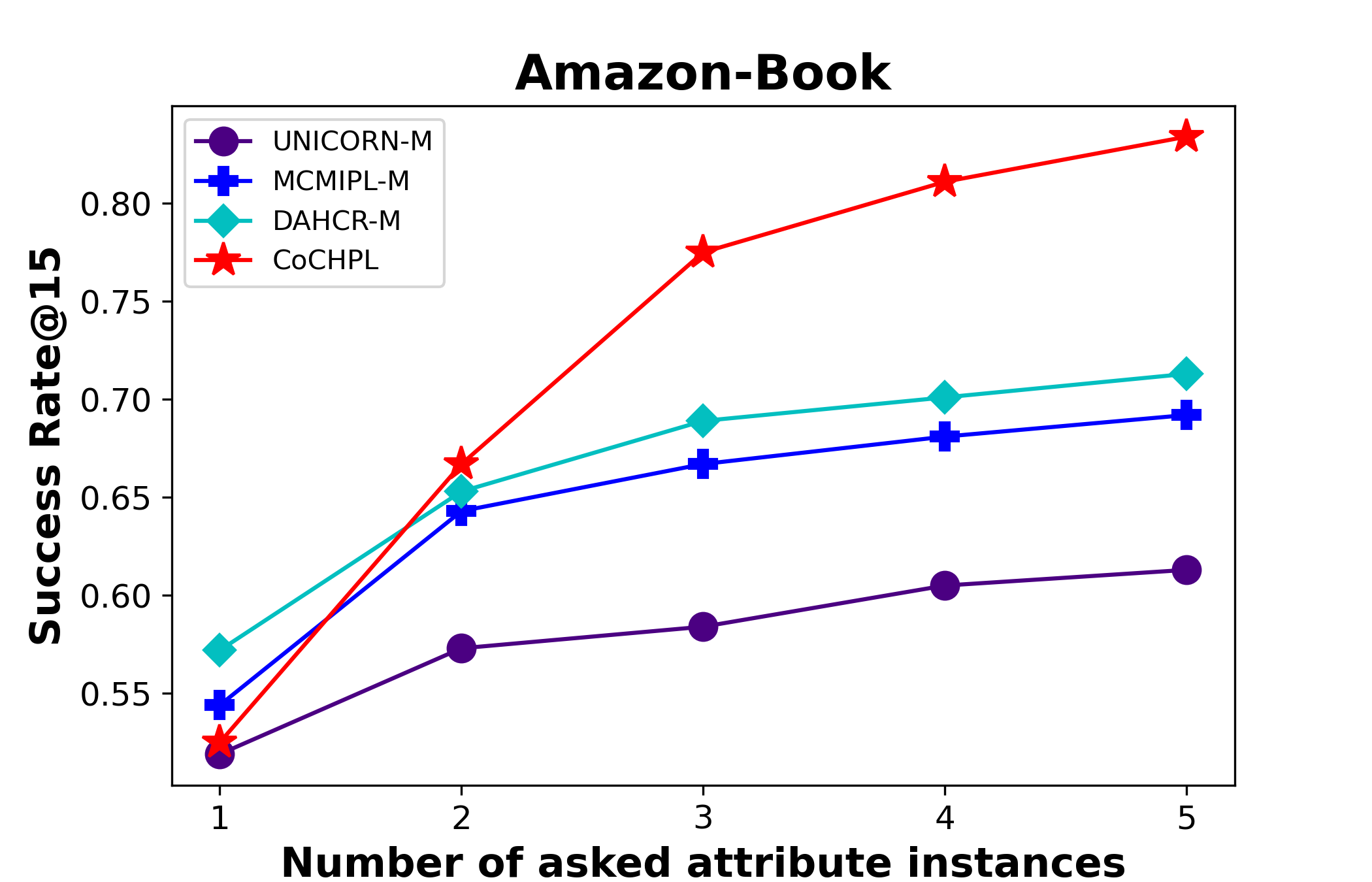}}
	\end{minipage}
 
	\caption{Performance comparisons of different asked attribute instances numbers in Yelp and Amazon-Book datasets.}
	\label{fig:attribute_instance_number}
\end{figure}
\subsubsection{Effect of Attribute Instances Number and Case Study (RQ3).}
\begin{figure}[t]
	
	\begin{minipage}{0.5\linewidth}
		\centerline{\includegraphics[width=\textwidth]{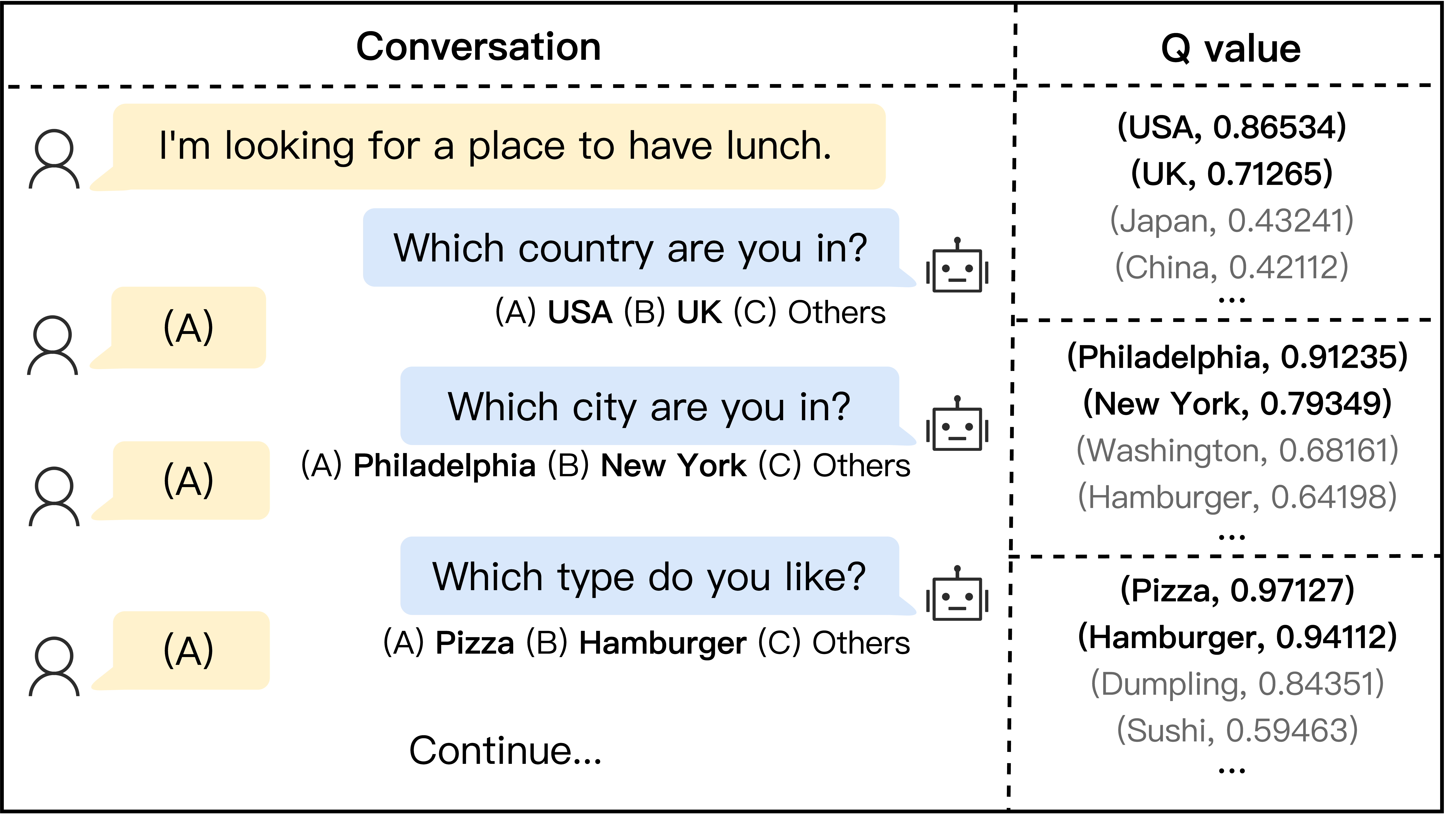}}
		\centerline{(a) DAHCR-M}
	\end{minipage}
	\begin{minipage}{0.5\linewidth}
		\centerline{\includegraphics[width=\textwidth]{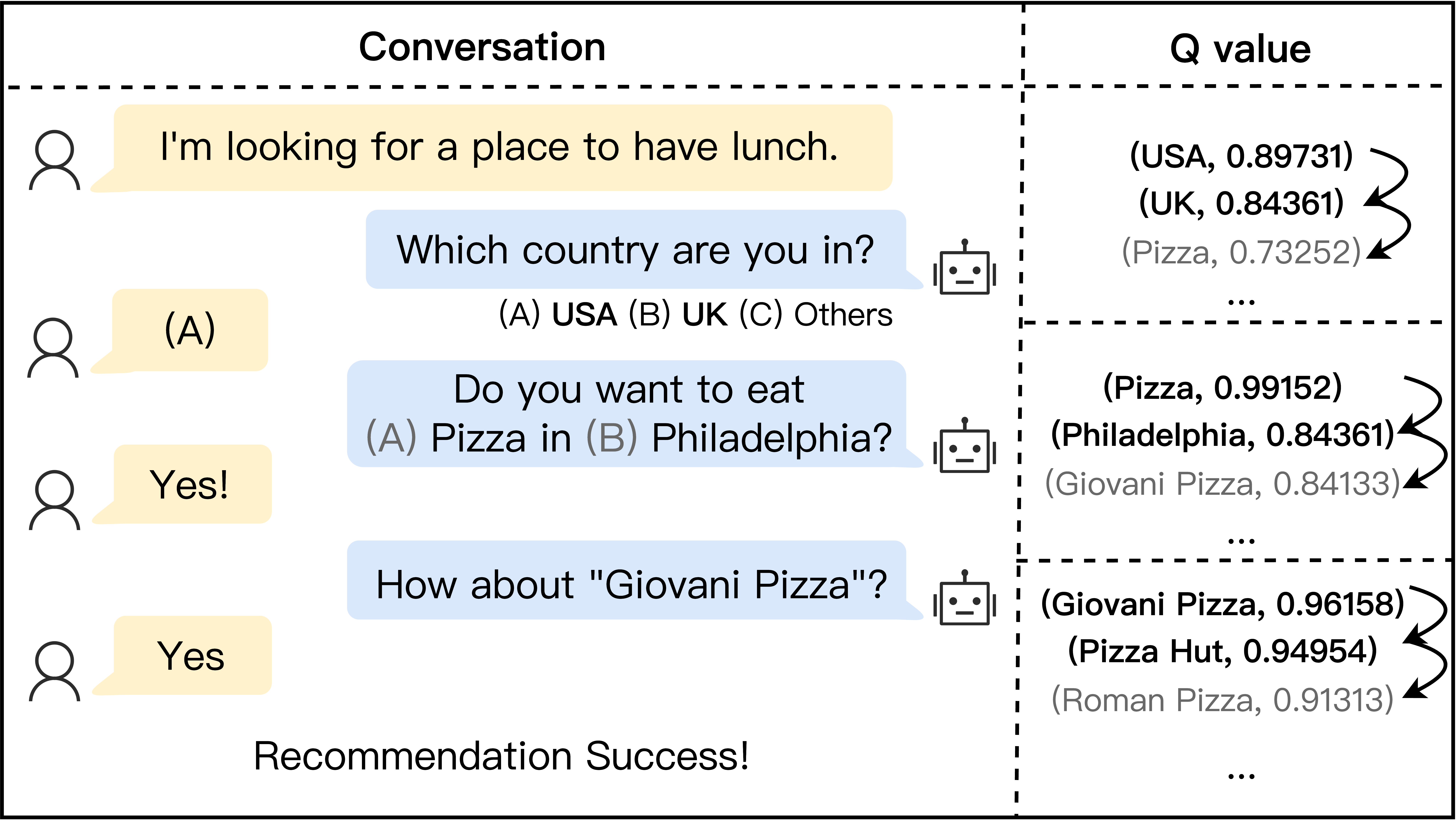}}
		\centerline{(b) CoCHPL}
	\end{minipage}
 
	\caption{Conversations generated by DAHCR-M and CoCHPL, and the process of choosing choices in each turn. By utilizing prediction and inference stages, CoCHPL is capable of generating a more coherent and logical sequence of choices.}
	\label{fig:case_study}
\end{figure}

Our study, using Yelp and Amazon-Book datasets, assessed the reasoning capabilities of our model, CoCHPL, against three baselines: UNICRON-M, MCMIPL-M, and DAHCR-M. We varied the number of attribute instances requested from 1 to 5 per turn as shown in \figref{fig:attribute_instance_number}.
We noted that while the success rate of baseline models improved with an increased number of attribute instances, the marginal gains diminished over time, approaching a plateau. This trend suggests that these models tend to select numerous but less effective attribute instances, leading to suboptimal exploitation of user preferences. In contrast, CoCHPL demonstrated a nearly linear performance improvement in both datasets. A case study depicted in \figref{fig:case_study} further reinforces CoCHPL's superiority. Unlike DAHCR, which selects the top-$2$ attribute instances based on Q-values, often resulting in homogenous choices due to similar embeddings, CoCHPL dynamically predicts user feedback and recalculates Q-values after each selection, prioritizing the top-$1$ attribute for subsequent choices. Both evaluations demonstrate that CoCHPL effectively diversifies and deepens the dependency among attributes, significantly enhancing the overall recommendation quality.

\section{Conclusion}

In this paper, we propose a more realistic CRS setting called Multi-Type-Attribute Multi-round Conversational Recommendation~(MTAMCR). To optimize the questioning efficiency, we develop a novel framework, namely Chain-of-Choice Hierarchical Policy Learning~(CoCHPL), for the MTAMCR scenario. 
Specifically, by formulating MTAMCR as a hierarchical RL task, we employed a long-term policy over action options to decide when to ask or recommend and two short-term intra-option policies to sequentially generate the optimal multi-type attribute and item chains with the assistance of the learnable termination and feedback prediction functions. 
Experimental results demonstrate that CoCHPL significantly outperforms state-of-the-art CRS methods in terms of both the reasoning ability in chains of choice and the success rate of recommendations.

\subsubsection{\ackname} This research was supported in part by the National Natural Science Foundation of China under Grant No. 62102110, 92370204, the Guangzhou Basic and Applied Basic Research Program under Grant No. 2024A04J3279, and the Education Bureau of Guangzhou Municipality.
%
%
%
\bibliographystyle{splncs04}
\bibliography{references}

\end{document}